\documentclass[twocolumn]{webofc}
\usepackage[varg]{txfonts}        
\usepackage[dvipsnames]{xcolor}

\usepackage{caption}

\title{Terzina on board NUSES: a pathfinder for EAS Cherenkov Light Detection from space}

\author{
    \firstname{Leonid}  \lastname{Burmistrov}  \inst{1} \fnsep\thanks {email:{leonid.burmistrov@unige.ch}}    ~for the NUSES Collaboration \\ (a complete list of authors can be found at the end of the proceedings) 
 } 
 
\institute{Département de Physique Nucléaire et Corpusculaire, 
  Faculté de Sciences,  Université de Genève, 1205, Switzerland }
\abstract{In this paper we introduce the Terzina telescope as a part of the NUSES space mission. This telescope aims to detect Ultra High Energy Cosmic Rays (UHECRs) through the Cherenkov light emission from the extensive air showers (EAS) that they create in the Earth's atmosphere. The Cherenkov photons are aligned along the shower axis inside about $\sim 0.2-1^{\circ}$, so that they become detectable by Terzina when it points towards the Earth's limb. A sun-synchronous orbit will allow the telescope to observe only the night side of the Earth's atmosphere. In this contribution, we focus on the description of the telescope detection goals, geometry, optical design and its photon detection camera composed of Silicon Photo-Multipliers (SiPMs). Moreover, we describe the full Monte Carlo simulation chain developed to estimate Terzina's performance for UHECR detection. The estimate of the radiation damage and light background rates, the readout electronics and trigger logic are briefly described. Terzina will be able to study the potential for future physics missions devoted to UHECR detection and to UHE neutrino astronomy. It is a pathfinder for missions like POEMMA or future constellations of similar satellites to NUSES.}
\begin{document}

\maketitle

\section{The NUSES mission and its scientific goals}
\label{intro}

\begin{figure}[ht]
  \centering
  \includegraphics[width=7cm,clip]{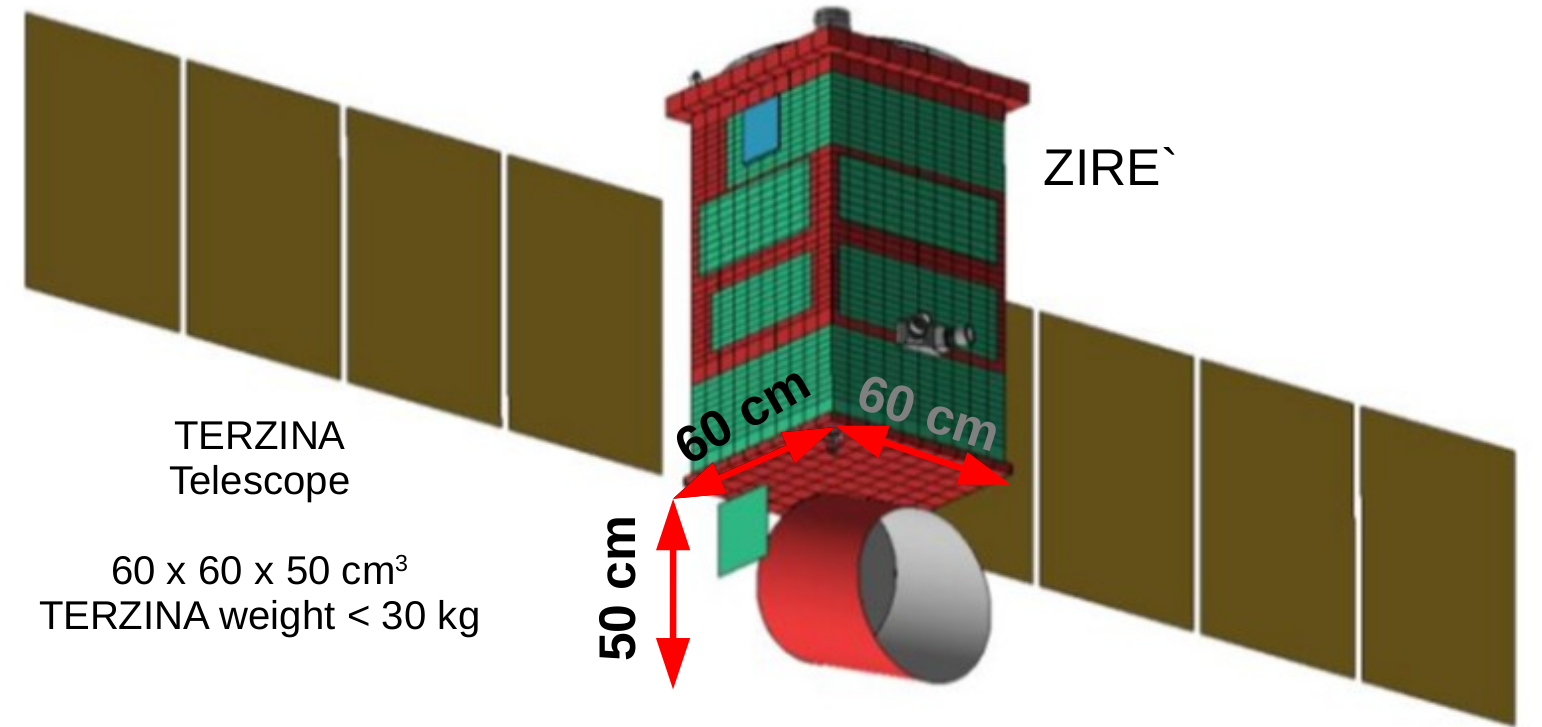}
  \caption{The NUSES satellite containing the Ziré apparatus at the top and the Terzina telescope at the bottom during the flight.}
  \label{fig:nuses}
\end{figure}

NUSES is a pathfinder satellite project (Fig.~\ref{fig:nuses}) for two innovative detectors dedicated to the study of cosmic radiation and the Sun-Earth environment \cite{ref:tocome}. They are Terzina, devoted to high-energy cosmic rays beyond 100~PeV, and Ziré devoted to cosmic rays below 250~MeV. The NUSES mission will last about three years and the satellite, with the two payloads, will orbit at 550~km altitude at the beginning of life (BoL) and at about 525~km at the end of life (EoL). The orbit inclination will be $97.6^{\circ}$ with LTAN 18:00:00.

\begin{figure*}
  \centering
  \includegraphics[width=11cm,clip]{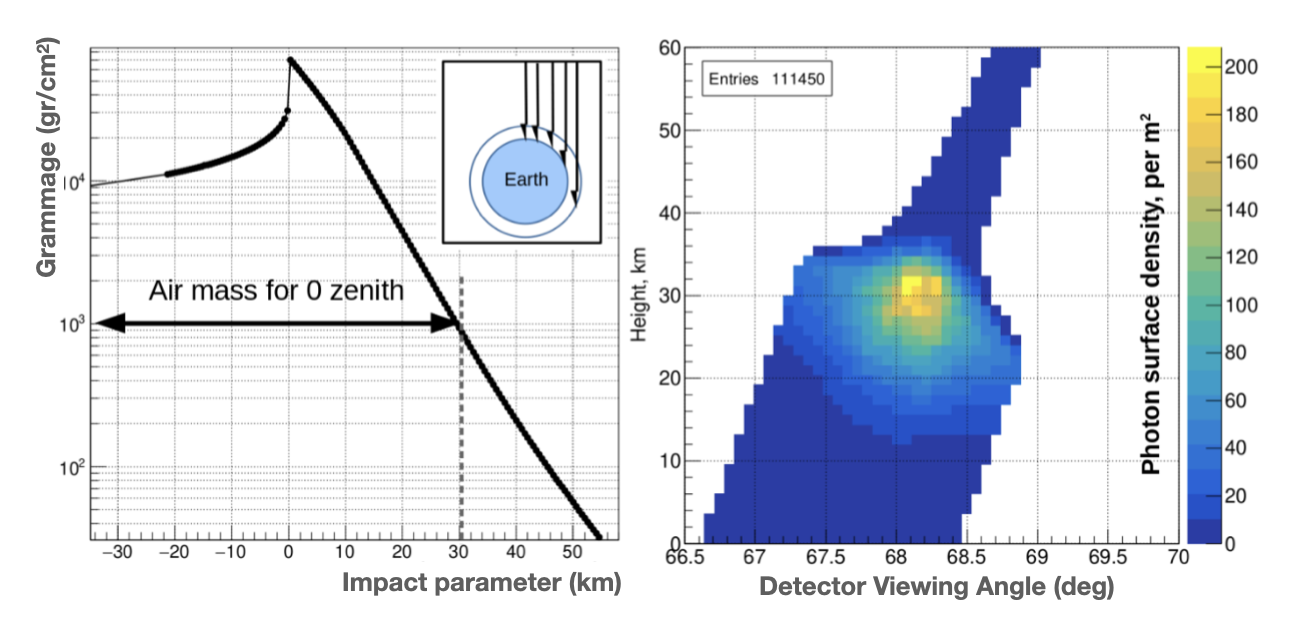}
  \caption{\small{{\bf Left:} Grammage of a cosmic particle as a function of its impact parameter. The insert at the top-right schematically shows incoming particles with different impact parameters. In case the impact parameter is larger than the radius of the Earth, the maximum air mass seen by the particle doubles. {\bf Right :} Photon surface density (z-axis) as a function of the viewing angle and height of a UHECR (primary particle: 100~PeV proton) estimated with EASCherSim~\cite{pub:easchersim} (see Sec.~\ref{sec:simulation}). The total number of simulated protons is about $1.1 \times 10^5$.}}
  \label{fig:protons}
\end{figure*}

Ziré is made of a scintillating fiber tracker, a stack of plastic scintillator counters and an array of LYSO crystals. An active veto system and a Low Energy Module (LEM) are also part of the payload. SiPM will be used as light sensors. The instrument will perform spectral measurements of electrons, protons and light nuclei from few up to hundreds of MeV, also testing new tools for the detection of 0.1-10~MeV photons~\cite{Adriano:2022}, and monitoring Van Allen radiation belts and space weather effects. 

\begin{figure*}
\centering
\hfill
\begin{minipage}[b]{\columnwidth}
  \centering
  
  \includegraphics[width=8cm,clip]{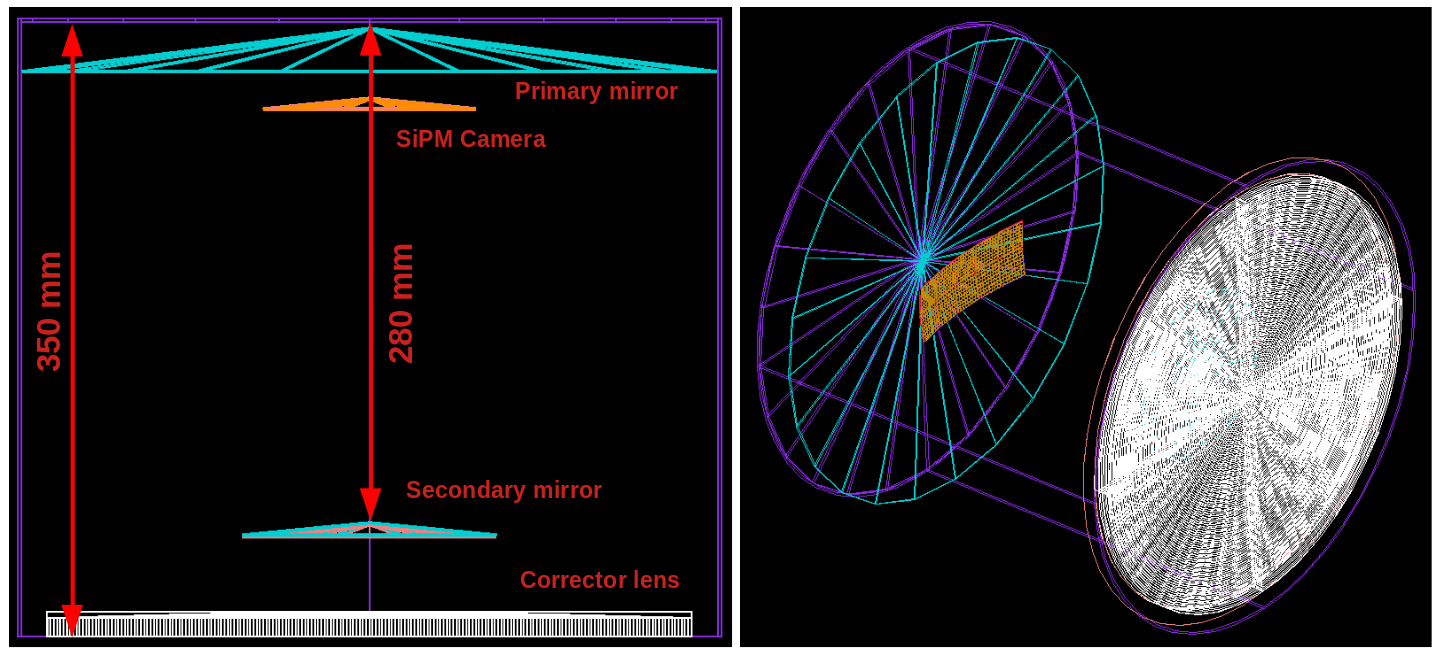}
  \caption{\small{{\bf Left :} View from the top of the payload of Terzina. {\bf Right :} Side view and SiPM camera.}}
  \label{fig:optics}
\end{minipage}\hfill
\begin{minipage}[b]{\columnwidth}
  \centering
  \begin{tabular}{|l|l|l|l|}
    \hline
                 & RoC$^{*}$  & Distance           & Diameter \\
                 &      & to primary mirror      &          \\
    \hline
    units        & m    & mm                 & mm       \\
    \hline
    Big mirror   & 0.80 & 0   & 394 \\
    Small mirror & 0.36 & 280 & 144 \\
    Camera plane & 0.30 & 40  & 121 \\
    Corrector    &  -   & 350 & 362 \\    
    \hline
        \multicolumn{4}{|l|}{Equivalent focal length : 925 mm} \\
    \hline
    \hline
        \multicolumn{4}{|l|}{$^{*}$RoC - Radius of Curvature} \\
    \hline
  \end{tabular}

  \caption*{Table 1: Parameters of the optical system}\label{tab:optics}
  \end{minipage}\hspace*{\fill}
\end{figure*}

Terzina is a satellite-based detector designed for Cherenkov light detection from EAS induced by UHECR in the Earth's atmosphere. Terzina will aim to the detection of Earth-skimming ultra-high energy neutrinos \cite{Cummings:2020ycz} and ultra-high energy cosmic rays with energy above 100~PeV \cite{pub:main_pub}. It will prove for the first time the space-based detection of Cherenkov light from EAS. The detection principle is discussed in~\cite{pub:main_pub,pub:poemma,pub:poemma_pub}.

We expect to see the highest rate of detectable signal from UHECRs when their interactions happen at the atmospheric height of about 30~km for protons of 100~PeV (see~Fig.~\ref{fig:protons})

\section{The Terzina telescope}
\label{sec:terztel}
The Terzina payload is composed by the following elements: the optical head unit, which is a near-UV-optical telescope, the focal plane assembly (FPA), the thermal control system, and the external harness and electronic units which will be in a separate box. 

The optical system of the telescope is based on a dual mirror configuration composed of two parabolic primary and secondary mirrors with a corrector lens in order to cope with  aberrations on the photon detection plane, namely the FPA. The dual mirror configuration is chosen to maximize the focal length in the available space which is a envelope in the shape of a cut-cone with a 394~mm diameter and a 350~mm length. The resulting focal length is about 925~mm. The telescope is inclined by $67.5^{\mathrm{o}}$ with respect to nadir, having an optical axis pointing towards the Earth's limb (see Figures~\ref{fig:nuses} and \ref{fig:optics}). 

The FPA is conceived to detect photons from below and above the limb. It has a rectangular shape with a $2:5$ aspect ratio. It is composed of 10 SiPM arrays of $8~\times~8$ pixels each and  the point spread function (PSF) of the optical system and the dimension of the limb from the orbit of Terzina are compatible with a $3~\times~3$~mm$^{2}$ pixel size forming 2 rows of 5 arrays each. The sensors are provided by the Fondazione Bruno Kessler (FBK) and described below. The telescope has a  field-of-view of $7.2^{\mathrm{o}}$ horizontally and $2.5^{\mathrm{o}}$, as each pixel sees  $0.18^{\mathrm{o}}$. It can observe a vast volume of the atmosphere with a cross-section of $140~\times~360$~km$^2$.

\begin{figure}
  \centering
  \includegraphics[width=7cm,clip]{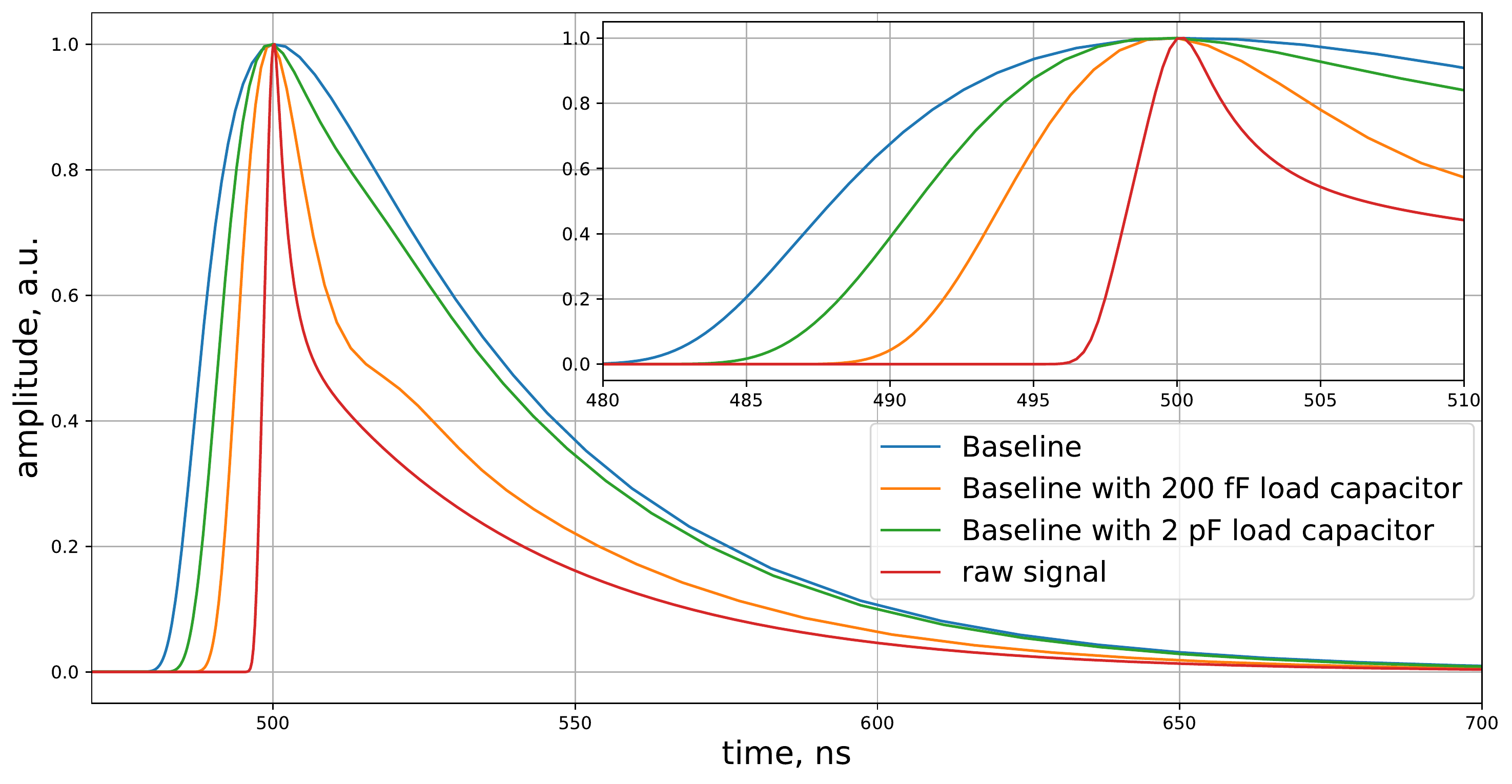}
  \caption{The SiPM signal shape before and after different amplifier configurations for the NUV-HD-MT from FBK with 25~$\mu$m micro-cell size. The amplitudes are normalized to the maximum value.}
  \label{fig:signal}
\end{figure}

The camera frontend electronics is composed of 10 Application Specific Integrated Circuits (ASICs), each reading out one SiPM module with 64 channels. The ASIC has an input amplification stage and only digitizes signals upon validation of trigger conditions described in Sec.~\ref{sec:trigger}. 
The currently foreseen amplifier has a bandwidth of 35~MHz forming the amplified signal with $\sim10$~ns rising time. In Fig.~\ref{fig:signal} different scenarios are reported. The red line is the raw signal as provided to the input of the preamplifier. The blue line is the output of the preamplifier with 2~pF output load. After some optimizations to the bias parameters the preamplifier response was improved in terms of speed. The power consumption of the stage was maintained constant, while its noise performance was reduced by a factor 2. Nonetheless the noise increase is tolerable, due to the amplitude of signals to be detected. The orange and green lines represent the preamplifier output with capacitive loads of respectively 200~fF and 2~pF at its output. The 200~fF value is more realistic while the 2~pF is a worst case scenario.
The ASIC samples the analog signals at a frequency of 200~MHz. It will digitize at least 3 points on the rising edge of the signal.
The digitized signals from the ASICs will be brought out of the telescope system through a hole in the primary mirror partly obscured by the secondary, to a box where the electronic harness is located. The box will include an FPGA collecting the data from all 10 ASICs to form the trigger described in Sec.~\ref{sec:trigger}.
The full Terzina SiPM camera plane with the ASIC system should not consume more than 5 W.

Due to the large NGB and dark current rate (DCR), the sensor technology has to be chosen taking into account many factors. Among the others described in the paragraph below, it is important to take into consideration the power consumption and the radiation tolerance.

Terzina has a technological driver to explore the performance of silicon photomultipliers (SiPMs) in space. Compared to the
classic photomultipliers, they are preferred for their typically higher Photon Detection Efficiency (PDE) and their robustness, although they suffer a larger background due to the higher sensitivity, and sensitivity to radiation damage. We developed a parametric waveform simulation based on the knowledge of the sensor's single photoelectron signal shaped by the amplifier and the sensor's noise rate and NGB. This simulation, together with the Geant 4 full simulation, is useful to understand the requirements of the relevant parameters of the sensor and also to define the trigger (see Sec.~\ref{sec:trigger}). Moreover, the following requirements for the SiPM operating properties are needed: the PDE at peak wavelength should be at least of 50\%, the direct and delayed cross-talk CT should be lower than 10\% at operation voltage, as well as afterpulse AP and the DCR preferably not higher than 100~kHz/mm$^2$.


The NUV-HD-LowCT SiPM technology~\cite{pub:NuV-HD}, developed by FBK, has typical values (for 35~$\mu$m cell-size) of dark count rate (DCR $\sim$ 100 kHz/mm$^{2}$), afterpulsing (AP $\sim$ 5\%;) and optical crosstalk (CT $\sim$ 5\%–20\%;), and photodetection efficiency (peak PDE $\sim$ 50\%–60\%.) in the blue region of the light spectrum. The highest values of the PDE are achieved with the largest cell sizes available (e.g. 35 – 40~$\mu$m), as they feature the highest fill factor (i.e. ratio between active area and total area of one micro-cell / SPAD). On the other hand,  the recovery time of the micro-cell also increases with increasing cell size, as it is proportional to the micro-cell capacitance and, as a first approximation, to its area. 

To further improve performance, the Collaboration is also evaluating an upgraded version of the NUV-HD technology, the NUV-HD-MT, which employs metal filling of the Deep Trench Isolation (DTI) that separates adjacent micro-cells in the SiPM, to further reduce optical cross-talk probability without sacrificing PDE. 

It is to be noted that one of the expected background in space is due to relativistic charged particles which can produce Cherenkov light in the telescope optical/mechanical parts and the SiPM protective resin layer. The potential background created by cosmic ray protons and, to a reduced extent, by relativistic heavier nuclei~\footnote{These are mostly He with lower abundance of heavier nuclei up to Fe, particularly dangerous since number of Cherenkov photons is squarely scales with particle charge.} and electrons is mitigated by reducing these elements close to the sensors and by the telescope chassis with baffle.  
 
\begin{figure}
  \centering
  \includegraphics[width=8cm,clip]{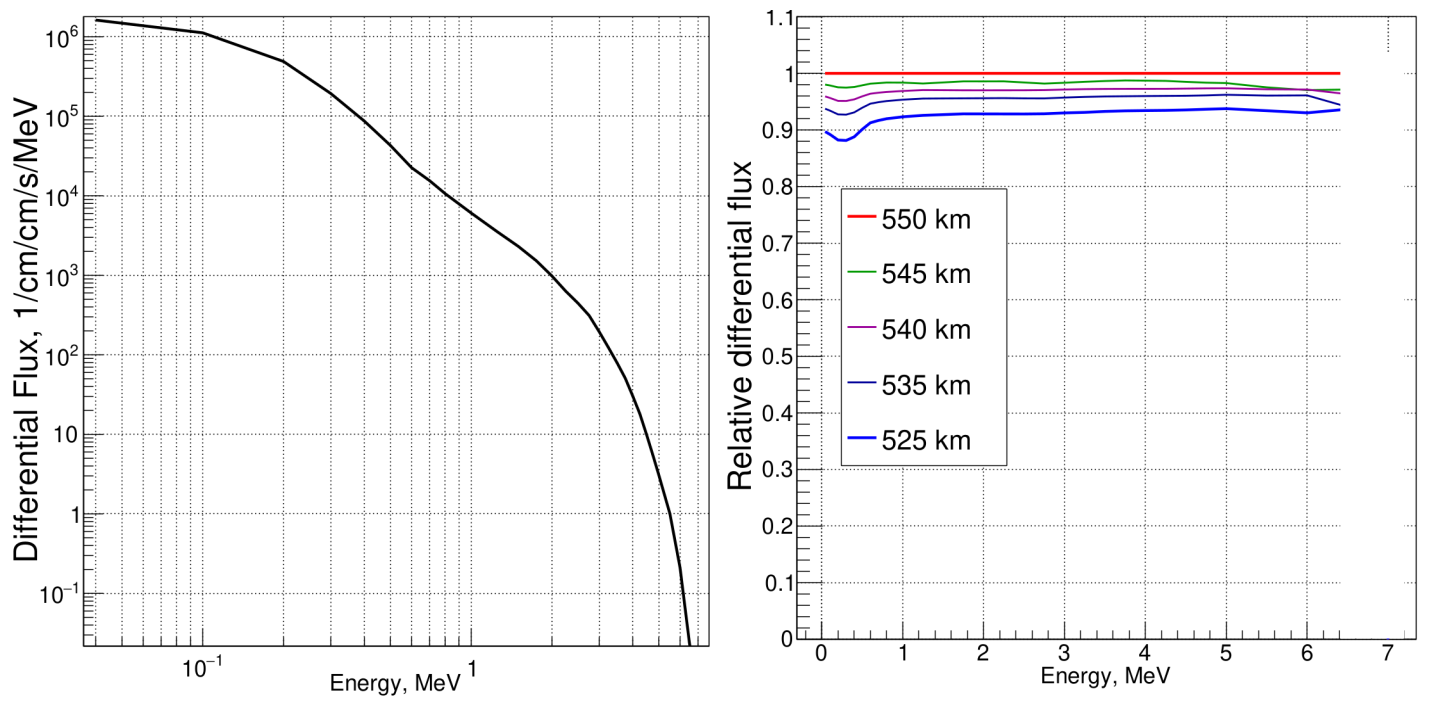}
  \caption{\small{{\bf Left:} Average differential electron flux at 550~km height obtained with SPENVIS~\cite{pub:SPENVIS}. {\bf Right:} Relative variation of the differential electrons flux for different heights.}}
  \label{fig:ele_bg}
\end{figure}

\begin{figure}
  \centering
  \includegraphics[width=8cm,clip]{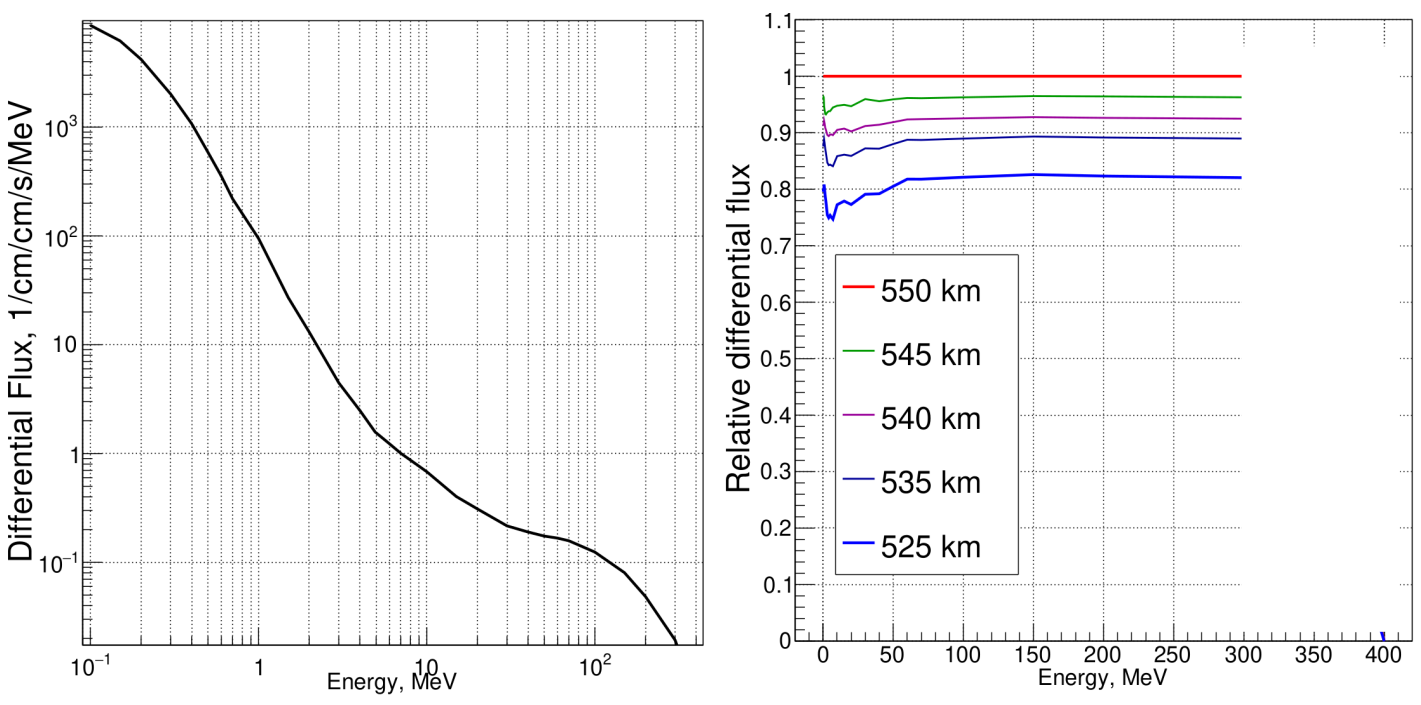}
  \caption{\small{{\bf Left:} Average differential proton flux at 550~km height (SPENVIS~\cite{pub:SPENVIS}). {\bf Right:} Relative variation of the differential proton flux for different heights.}}
  \label{fig:prot_bg}
\end{figure}

We expect electrons and protons to be the dominant sources of background. Ionizing radiation, like protons and heavier nuclei, induces damage in the silicon of SiPMs. Moreover, neutrons can create dislocations, within the crystal structure, increasing the DCR.
Electrons, on the other hand, produce secondary gammas causing non-ionising radiation damage.
The level of damage depends on the cosmic ray radiation spectra that we obtain with SPENVIS for Terzina's orbit~\cite{pub:SPENVIS} (see Fig.~\ref{fig:ele_bg} and Fig.\ref{fig:prot_bg}). We use the full simulation of the telescope to estimate the radiation dose integrated by the SiPMs knowing the electron and proton spectra (see sec.~\ref{sec:simulation}) . 

\begin{figure}
  \centering
  \includegraphics[width=8cm,clip]{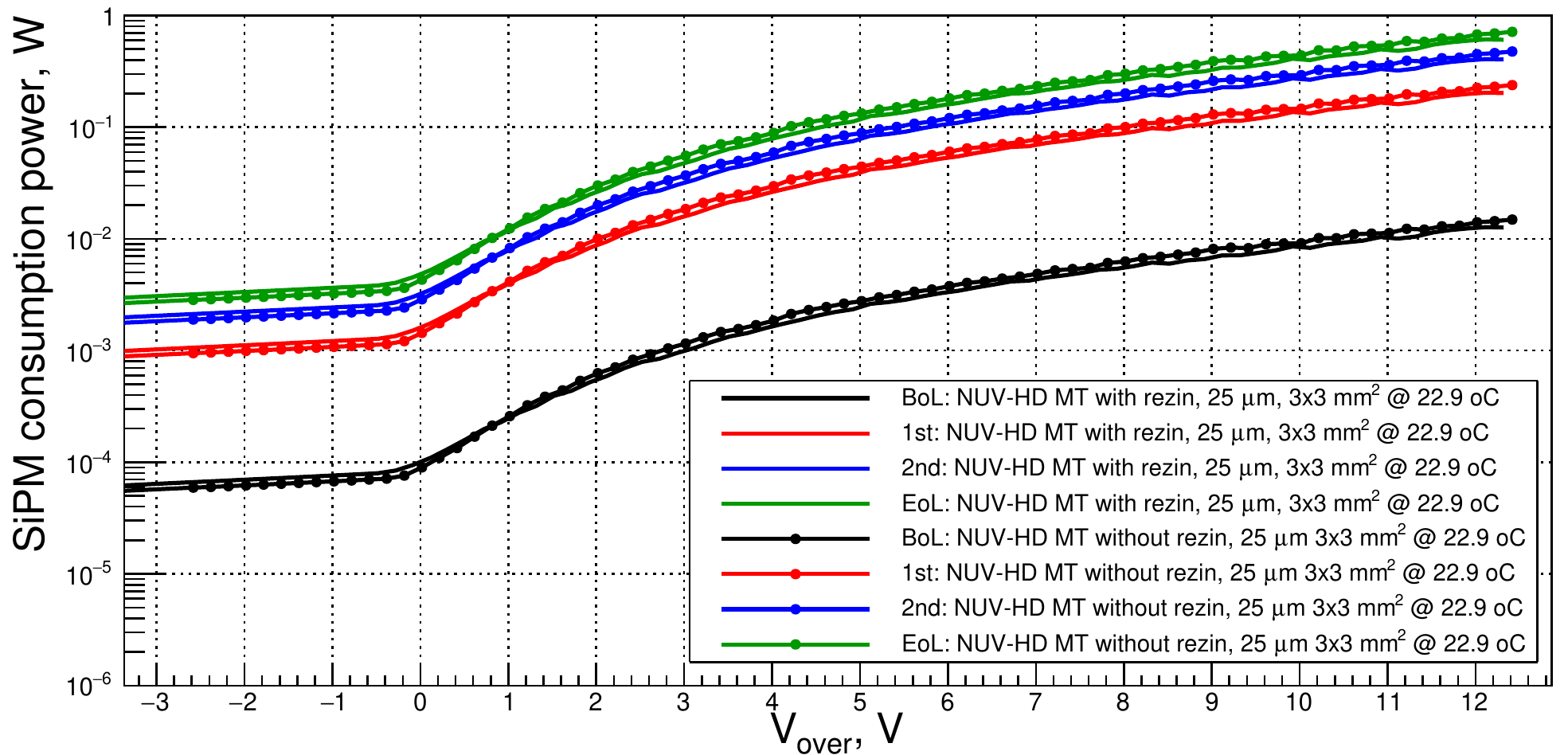}
  \caption{Power consumption by the SiPM camera.}
  \label{fig:camera_sipm_power}
\end{figure}

\begin{figure*}[h!]
  \centering
  \includegraphics[width=13cm,clip]{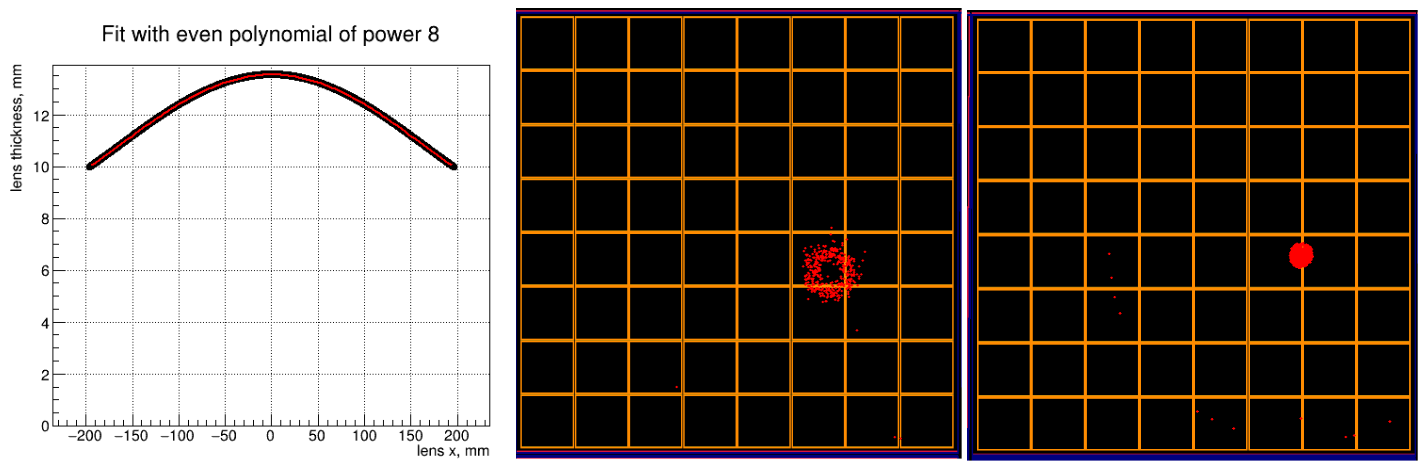}
  \caption{\small{{\bf Left:} The cross-section view of the corrector lens. {\bf Center:} Photons  induced by many UHECRs from 100~PeV protons, with equal parameters, superimposed on the camera plane. {\bf Right:} PSF at the same location on the camera plane.}}
  \label{fig:shower_im}
\end{figure*}

\begin{figure}[t!]
\centering
  \includegraphics[width=7.5cm,clip]{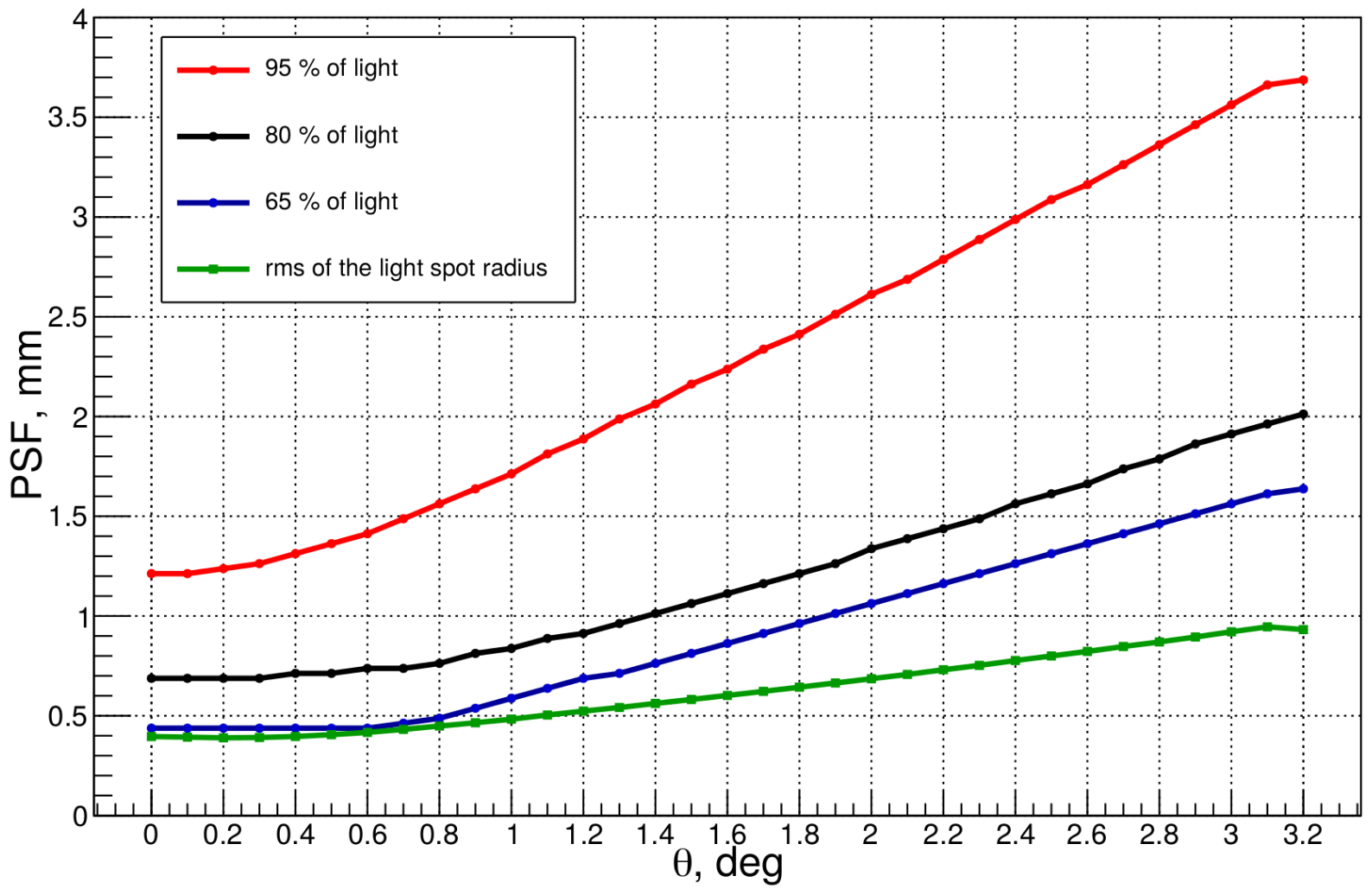}
  \caption{\small{RMS of the radius of the light spot before (lower curve) and after (3 upper curves for different percentage of photon containment) it is processed by the optical elements of the telescope from Geant 4 simulation. For brevity we call this PSF on the y-axis.}}
  \label{fig:psf_rms}
\end{figure}

In Fig.~\ref{fig:camera_sipm_power} the SiPM camera power consumption at different stages of the lifetime of the mission are shown. The black curve is measured in a laboratory test bench while the others are calculated using the linear extrapolation of the coefficient of the DCR increase with integrated radiation dose. Next, we can estimate NGB + DCR at BoL, after the first and second years of operation, and at EoL, namely for background levels of: 11~MHz, 22~MHz, 33~MHz, 44~MHz respectively. 
At the end of life, the power consumption of the sensors of the camera, operated at an over-voltage of about 6~V, will reach 0.2~W. This figure does not include the power consumption of the 10 ASICs, which are expected to consume 3.2~W reading out the 640 channels of the camera, and the trigger system for which the power consumption is estimated to be about 8~W.

Given all considerations and the fact that the NuV-HD-MT tests are not completed yet, our baseline solution is the NUV-HD-lowCT solution of FBK with
25~$\mu$m micro-cell size~\cite{pub:NuV-HD} and the optimal solution is NuV-HD-MT once validated with similar micro-cell size. If tuning of quenching resistor leads to a shorter slow component we can use 35~$\mu$m as well, ensuring larger PDE.
Despite we are aware that the NuV-HD-MT can probably be run at higher over-voltage than what we assume in simulation,  thanks to its low crosstalk, in Sec.~\ref{sec:trigger} we used parameters at 6~V over-voltage of 50\% PDE, DCR of 77 kHz/mm$^2$,  CR+AP of about 1\% + 6.6\%, $\tau_s = 40$~ns.
\section{The Geant 4 simulation and the performance study}
\label{sec:simulation}

We have produced the preliminary simulation of Terzina based on Geant 4~\cite{pub:geant4,pub:geant4_cite}. While the camera simulation resembles the system we are building, the optical system is preliminary. Recently, a final optimization was conducted in collaboration with a specialized company and this final configuration needs to be implemented in Geant 4 in the next future. Nonetheless, 
 our results should not change substantially as the main optical parameters are similar between the two configurations.

The optical system of the two mirrors, currently implemented as spherical and not parabolic as they will be, is shown in Fig.~\ref{fig:optics} with the rectangular camera between mirrors and its parameters are provided in Table~1.
The profile of the corrector lens is calculated to minimize the PSF for on-axis photons (see Fig.~\ref{fig:shower_im}).

The optical system simulation takes into account the mirror and corrector lens reflectivity, transparency, the SiPM quantum efficiency and geometry of the photon sensitive camera. The resulting root mean square (RMS) of the light spot on the camera plane is shown in Fig.~\ref{fig:psf_rms}.

The FPA has been simulated using the geometrical shape of arrays and their appropriate fill factors in the micro-cells and in the arrays. 

\begin{figure}[h]
  \centering
  \includegraphics[width=7.5cm,clip]{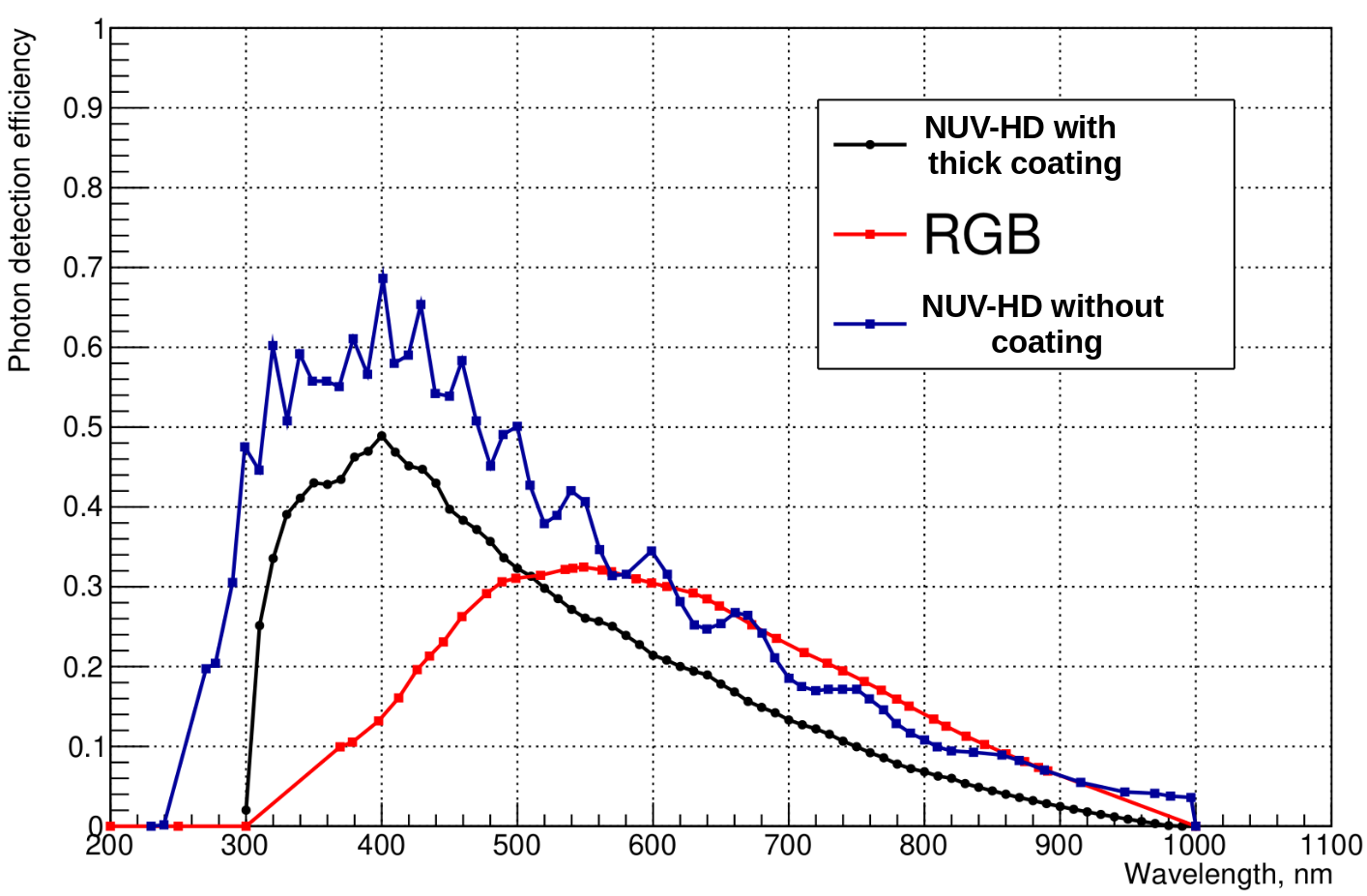}
 \caption{\small{Photon detection efficiency versus photon wavelength for different SiPM types by FBK.}}
  \label{fig:PDE}
\end{figure}

In the simulation we use the Photo-Detection Efficiency (PDE) of the FBK technology NUV-HD,
optimized for the UV band shown in Fig.~\ref{fig:PDE}, at an over-voltage of $\mathrm{V_{over}}$ = 6~V~\footnote{$\mathrm{V_{over}}$ =
$\mathrm{V_{operation}}$ (operational voltage) - $\mathrm{V_{bd}}$
(break down voltage)}. 
Also the PDE of a NUV-HD bare sensors
and of an RGB sensor (optimized for the optical-IR region) are shown. 

To estimate the sensitivity and performance to the expected UHECR signal, a dedicated generator was interfaced to the Geant 4 simulation.

\begin{figure*}[h]
  \centering
  \includegraphics[width=14cm,clip]{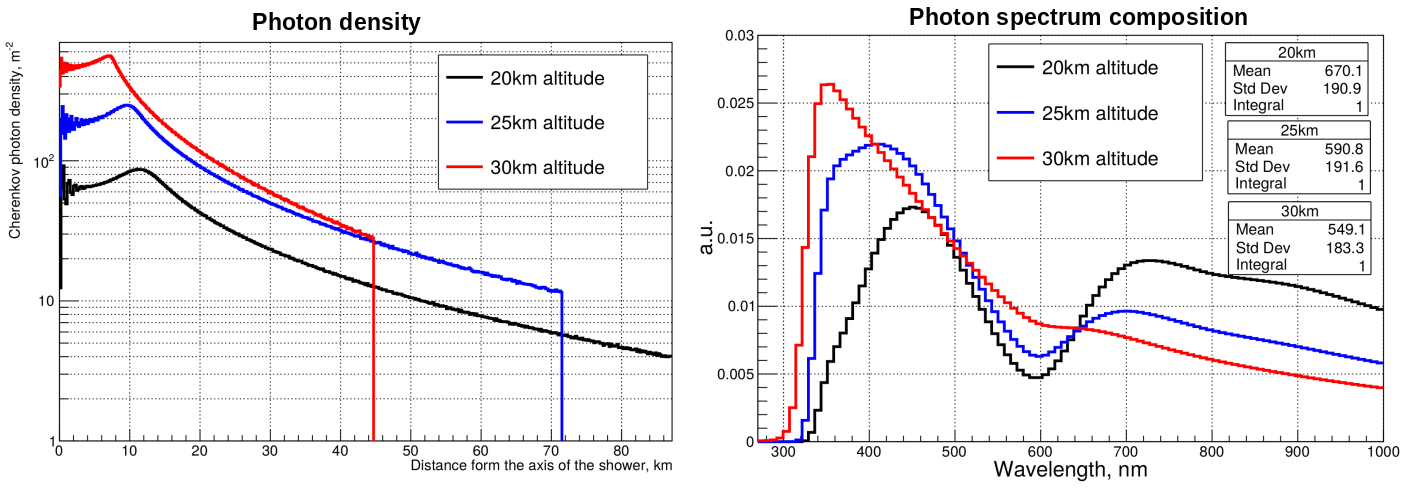}
  \caption{{\bf Left :} Photon density of the Cherenkov ring as seen from the telescope
  produced by 100~PeV protons for different heights as a function of distance from the shower axis. {\bf Right:} Photon spectral composition for showers produced at different altitudes in the atmosphere. The deeper the shower the more the photons will be scattered in the atmosphere and spectral emission moves toward the red band.  
  }
  \label{fig:photon_den}
\end{figure*}

\begin{figure*}[ht]
  \centering
  \includegraphics[width=14cm,clip]{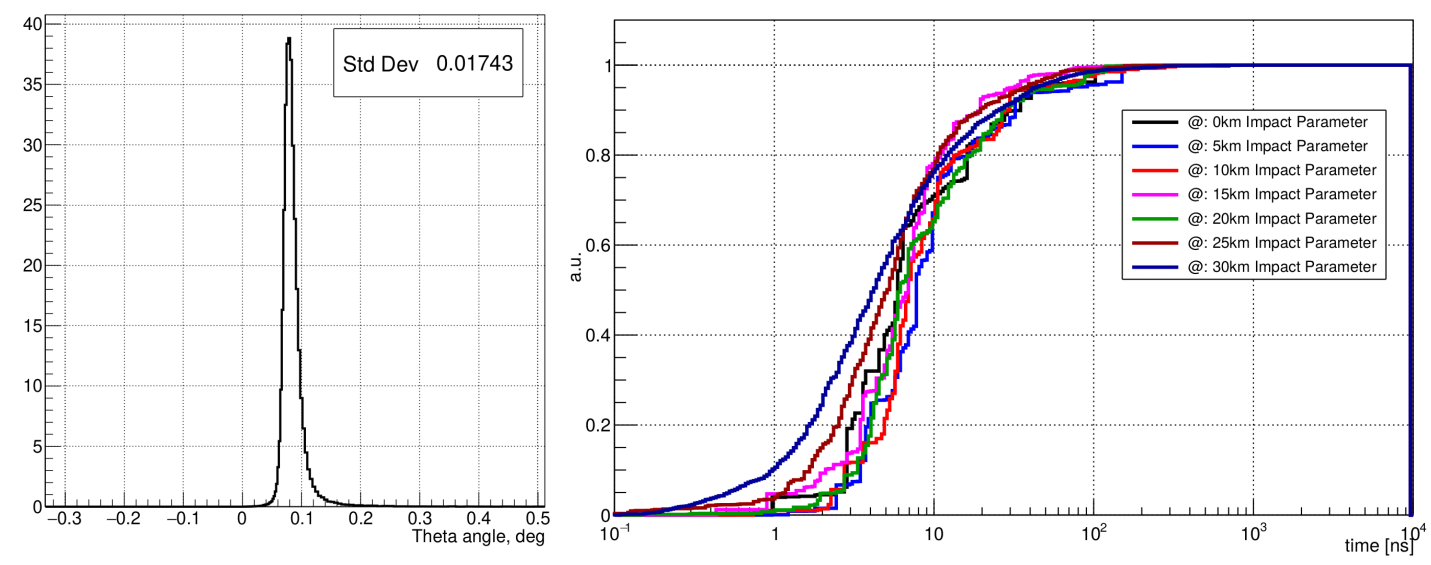}
  \caption{\small{{\bf Left :} Photon angle distribution for a proton of 100 PeV with respect to the shower axis seen by the telescope pointing in the direction of this axis. 
  {\bf Right :} Cumulative distribution of the Cherenkov photons as a function of time from the first one emitted in the direction of the telescope. 
  }}
  \label{fig:photon_time}
\end{figure*}

 We use the Emission for Extensive Air Showers Cherenkov Simulation (EASCherSim)~\cite{pub:easchersim} as physics event generator. EASCherSim provides the average photon density, photon spectral composition (Fig.~\ref{fig:photon_den}), the photon angle with respect to the shower axis and time (Fig.~\ref{fig:photon_time}) of photons. We assume a uniform distribution of the azimuth angle with respect to the telescope axis and the absence of the correlation between photon angles, time and wavelength. For the moment it can only simulate proton interactions. The composition of UHECRs is mixed at 100 PeV (see e.g.~\cite{icetop}). Additionally Pierre Auger found out that it is dominated by light elements at 2000 PeV \cite{PAO_2000} and becomes heavier at higher energies. Hence, in the future we will investigate a mixed composition.

An example of the shower image on the camera plane is shown in Fig.~\ref{fig:shower_im} (middle panel) next to the simulated PSF for the same location on the camera plane (right panel).

\begin{figure}[ht!]
  \centering
  \includegraphics[width=7cm,clip]{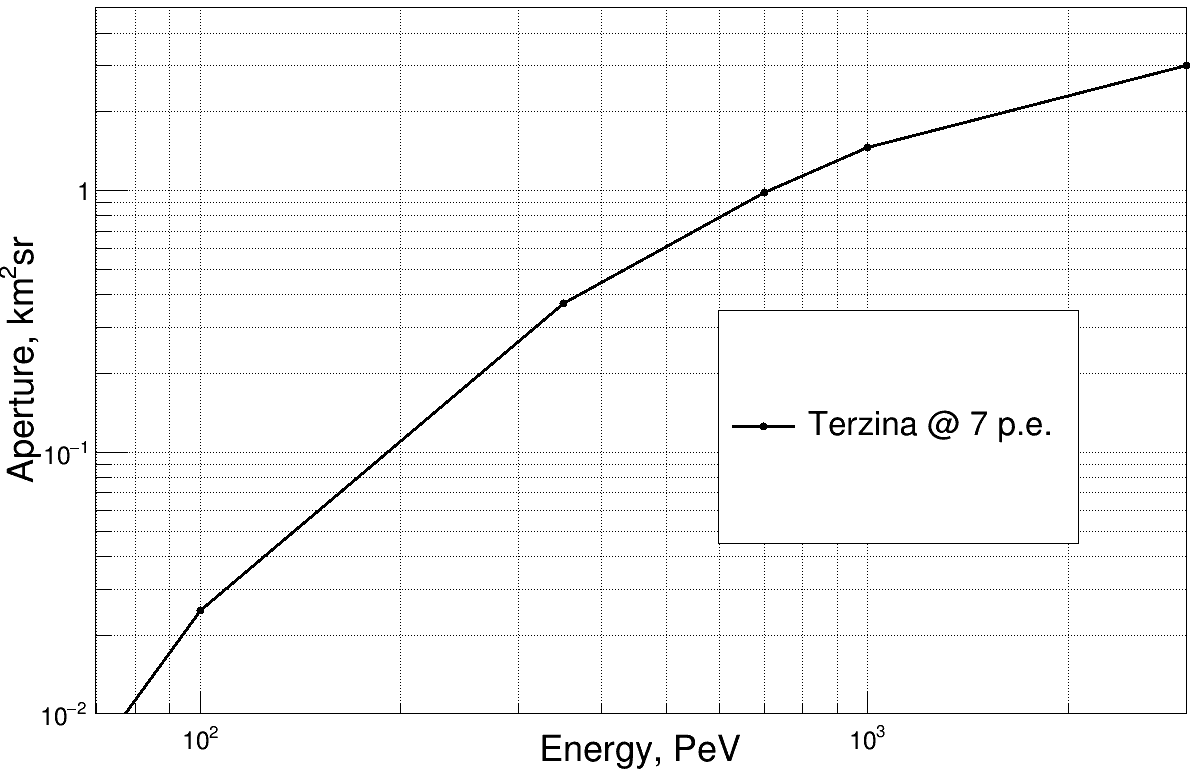}
  \caption{\small{The expected aperture for Terzina versus proton energy during the first year of operation.}}
  \label{fig:aperture}
\end{figure}
 
As an example of results on the Terzina performance from the simulation, we also show the aperture of Terzina. The plot assumes the NGB and DCR of the first year of operation (see Fig.~\ref{fig:aperture}). 
A threshold of 7 p.e.~per pixel, which suppresses the background, is assumed.


%

\section{The environmental background of Terzina: the night glow.}

The Night Glow Background (NGB) has been estimated in two different ways. Firstly, using the formula:

\begin{equation}
\label{eq:rate_NGB}
  \mathrm{Rate~per~pixel} = S \times d\Omega \times Flux \times Area \times PDE_{eff.} 
\end{equation}
where $ Area = 0.1~\mathrm{m^2}$; $PDE_{eff} = 0.1$ is the total optical efficiency calculated from the convolution of the SiPM PDE and the NGB spectrum, as a function of wavelength times an average optical of the 2 mirrors; $\Delta \Omega$ is the pixel solid angle; $\mathrm{Flux} [m^{-2}sr^{-1}ns^{-1}] \Big |^{\lambda=300~nm}_{\lambda=1000~nm} = 1.55 \times 10^4$; $S = 6$ is a conservative safety factor considering the possible largest variation of the glow, the rate per pixel is $\sim$10~MHz~\cite{pub:NGB}.

\begin{figure*}[h!]
  \centering
  \includegraphics[width=10cm,clip]{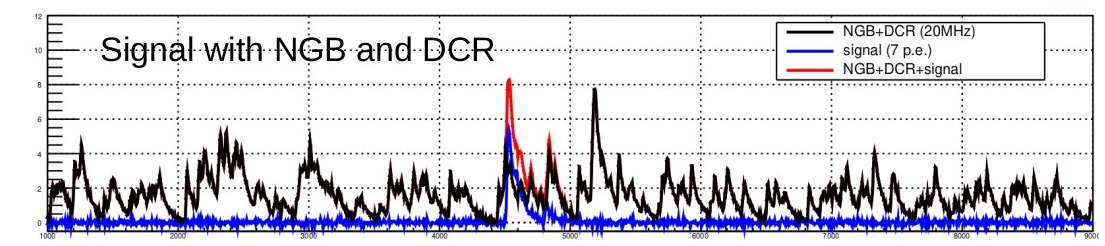}
  \caption{Signal with NGB and DCR}
  \label{fig:signalNGB}
\end{figure*}

Secondly, using the full simulation described in Sec.~\ref{sec:simulation}, we propagate photons assuming the NGB spectrum and a uniform angular distribution. Increasing the rate by a safety factor of 6 \cite{pub:poemma_pub}, which provide the typical variation with zenith angle of showers, we obtain a pixel background rate of $\sim$5~MHz. 
Given the discrepancy between the two estimates, we decided to use the more conservative value of 10~MHz per pixel (see Fig.~\ref{fig:signalNGB}).

\section{Event size and data throughput}
\label{sec:throughput}

The maximum daily data throughput for Terzina is 45~Gbit/day. Assuming that the data throughput will be dominated by camera events when compared to telemetry data, we can derive the maximum event rate.

One event will have the size of:
\begin{equation}
    \mathrm{size_{event}} = \mathrm{size_{event~header}} + \mathrm{N_{samples}}\times\mathrm{ADC_{resolution}}\times\mathrm{N_{pixels}}
    \label{eq:eventsize}
\end{equation}
where $\mathrm{size_{event~header}} = 34$~bits ( includes the bits dedicated to the alignment with the FPGA and and internal address of the readout units), $\mathrm{N_{samples}}= 20$, $\mathrm{ADC_{resolution}} = 12$~bit and the number of pixels, $\mathrm{N_{pixels}}= 9$.


At a trigger rate of 240~Hz, the maximum data throughput for Terzina is reached. From simulations we derive the thresholds to obtain 120~Hz of the events with signal contained in a single pixel and 120~Hz with signal shared between two neighboring pixels.

Within this work, we consider 9 pixels pad to readout. However, at the beginning of the mission, we will read one ASIC or even a full camera for debugging/calibration purposes. Latter as soon we get our system fully understood we will shrink the redout pad, dropping all irrelevant information.

\subsection{The trigger implementation}
\label{sec:trigger}

The ASIC offers the possibility of two programmable thresholds per channel hereinafter low and high threshold. The trigger will be structured so that, when the high threshold is passed by at least one channel, the waveforms of the 64 channels (one ASIC) are digitized and sent to the FPGA.

The time for the analog to digital conversion of all channels in the given ASIC, assuming the resolution of 12 bits takes $4096\times 5$~ns, namely  $\sim 20.4 \mu$s. The digitised data readout time is 64~(channels) $\times$ 32~(time-cells) $\times$ 12~(resolution) $\times$ 1.25~ns~(serialization in DDR) $= 30.7~\mu$s. Thus the total time for the analog to digital conversion and processing in the FPGA amounts to $\sim51.2~\mu$s.


Considering a power-law increase in the number of UHECR events with a decreasing energy threshold, in order to avoid an obvious increase of the background, the low threshold trigger is defined using the coincidence of 2 adjacent pixels. The coincidence time window is chosen to be twice the rise time of the SiPM signal after amplification, namely $\tau$ = 20~ns (however this time window is configurable).

 The implementation of the low level threshold trigger is the following: 
 when at least two pixels cross the low threshold in one ASIC or in the 16 pixels along the edges of two neighboring ASICs within the coincidence window $\tau$, a binary hit-map of the corresponding ASIC(s) is sent to the FPGA.
 The hit-map is analyzed to seek if there are at least two neighboring channels whose threshold is crossed (forming a cluster). If a cluster is found, the event is validated and the analog to digital conversion is performed, otherwise the event is discarded.

We decide to assign the same data rate to the two triggering mechanisms, namely 1/2 of the maximum rate each. In particular, considering the aforementioned data transmission limit, the rate per triggering mechanism is set to 120~Hz.
We assign the same rate to the two triggering mechanisms because the expected data rate of UHECRs hitting  1 or 2 pixels is about equal. Then, from simulation we evaluate the thresholds corresponding to this rate for both cases.
Regardless of whether the event is validated or discarded, the hit-map is used to increment counters to monitor the camera background.

 \subsection{Camera monitor}
The camera monitor based on hit-maps for low and high thresholds, aims to monitor the average background rate in the two parts of the camera:  the upper row of 5 arrays+ASICs in the camera plane will be sensitive to the Earth region and so to UHE neutrinos, but also to
undesired blind occurrences due to storms or city light pollution; the lower part of the camera plane will be sensitive to the limb. Very recent changes in the designing of the readout system opens the possibility to store on the FPGA the hit-map after the high-threshold trigger is satisfied. The single pixel hit-map is less vulnerable, with respect to double pixel, to the single point failure of a frozen combinatorial logic in the ASIC. This monitor has to operate constantly over mission duration. The monitor will deliver the rate at a given threshold for each pixel. The estimated monitoring data size, running at 1~Hz, is about 1-2\% of the event data rate. Hopefully, this method will allow to discriminate the human-induced or meteorological  spurious background also through correlations with meteo-maps. 


\begin{figure}
  \centering
  \includegraphics[width=6.5cm,clip]{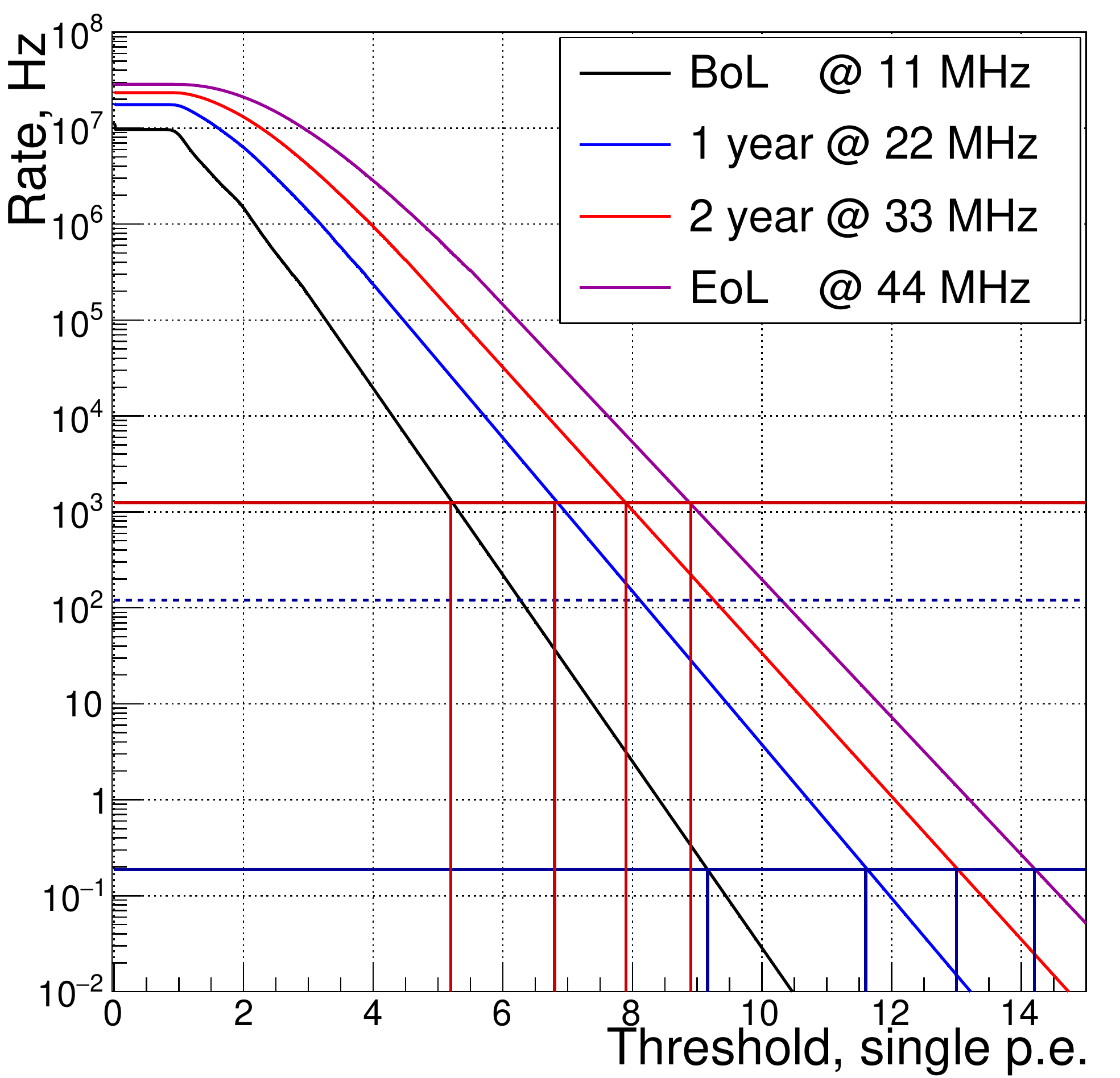}
  \caption{Single pixel trigger rate as a function of the threshold expressed in p.e. for DCR + NGB values estimated at different times during the mission life. The horizontal blue dashed line corresponds to the maximum trigger rate of (120~Hz). Horizontal blue line corresponds to the maximum single pixel trigger rate (120~Hz/640 $\sim$ 0.18~Hz per pixel). The horizontal red line corresponds to 1250~Hz (maximum single pixel rate for configuration with two pixels cluster). The vertical lines shows thresholds for single (blue) and double coincidence (red) trigger logic. Thresholds can be reconfigured during flight.}
  \label{fig:rate_vs_threshold}
\end{figure}

\subsection{Validation of the trigger through simulation}
\label{sec:valid_trig}

The parametric waveform simulation was used also to understand the requirements of the sensor and also where to set the trigger thresholds of the two configurations in order to satisfy the total data throughput limit of 45~Gbit/d. We simulate waveforms for a time frame of $10^7$~ns for different NGB+DCR conditions. Counting the numbers of peaks below a given threshold divided by the simulated time frame, we can estimate the trigger rates (see Fig.~\ref{fig:rate_vs_threshold}). 
For the results shown here, we assume the variable NGB during mission lifetime described in Sec.~\ref{sec:throughput} and a cross-talk of sensors of 6.6\% and DCR of 77 kHz/mm$^2$, as expected for NUV-HD-MT. 

For the single pixel trigger, which requires crossing the threshold in a single pixel, the threshold will have to be increased during the lifetime of the mission due to the increasing radiation damage as discussed in Sec.~\ref{sec:terztel}.
To satisfy the data throughput requirement, the threshold has to be set at the levels of:  9.2 p.e., 11.7 p.e, 13.1 p.e., 14.3 p.e (blue horizontal line), correspondingly at BoL, after the first year, after the second year, at EoL, respectively.

\begin{figure}[h]
  \centering
  \includegraphics[width=7cm,clip]{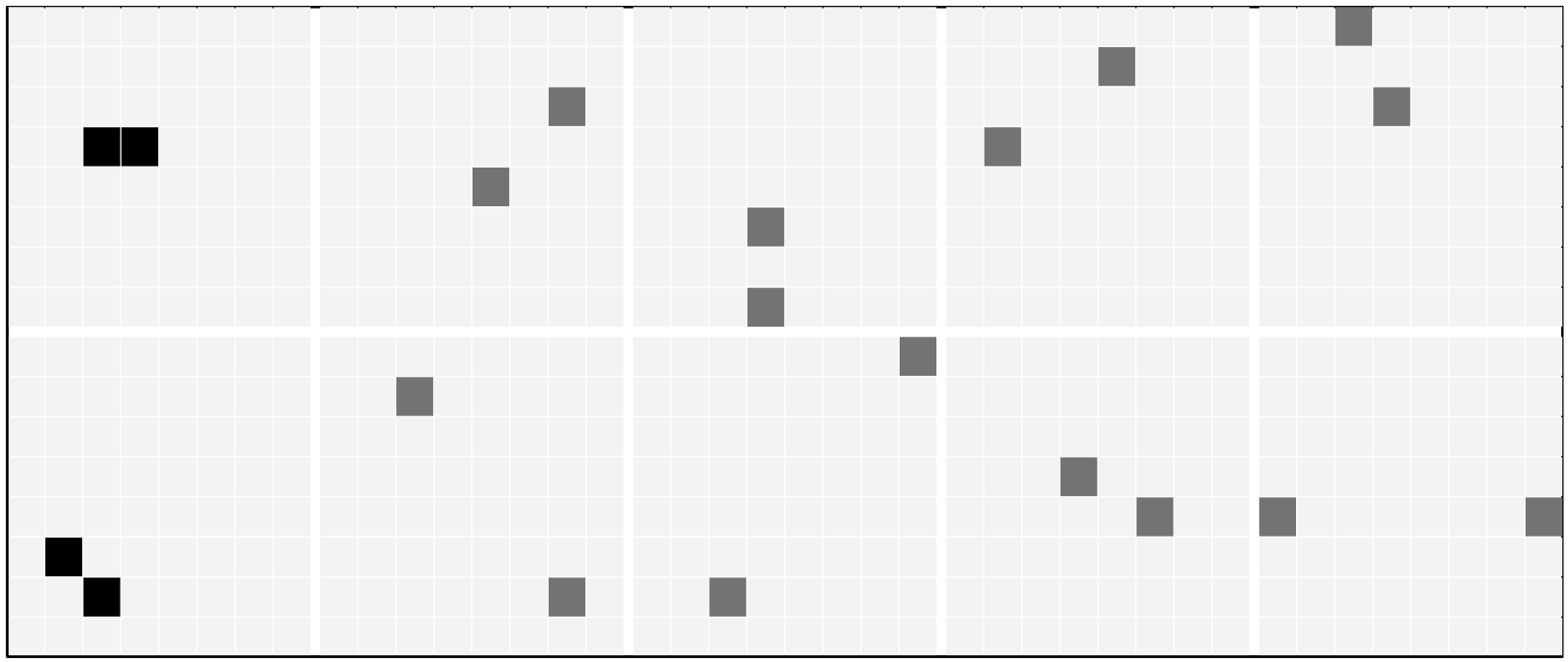}
  \caption{Camera composed of 10 arrays of SIPMs. Each array is readout by one ASIC with 8 $\times$ 8 channels. On the left, pixels in black forming a clusters can be seen, while the other fired pixels represent examples of background.}
  \label{fig:cluster_background}
\end{figure}

The number of possible pixel configurations which form a two-pixel cluster (see Fig.~\ref{fig:cluster_background}) in a single ASIC of 64 channels is equal to 210 or 2100 in the full camera. The number of all possible two-pixel configurations in a single ASIC is given by the $n-k$ combination equation:

\begin{equation}
    \frac{n!}{k!(n-k)!}\bigg\vert^{n=64}_{k=2},
\end{equation}

which yields a total of 2016 combinations per ASIC or 20160 in the full camera.
Considering the single pixel rate $R_1$ and the already defined concidence window $\tau$, the coincidence rate for clusters in the full camera is computed as follows:
\begin{equation}
    2 \times R_{1}^{2} \times \tau \times 2100.
\end{equation}

On the other hand, the coincidence rate for every possible combination of two pixels firing per ASIC in the full camera is:

\begin{equation}
    2 \times R_{1}^{2} \times \tau \times 20160 .
\end{equation}

The maximum single pixel rate for the coincidence configuration that can be afforded, given the data transmission rate, is therefore 1250~Hz (with fixed size of the readout pad - 9 pixels). Hence, from Fig.~\ref{fig:rate_vs_threshold} (red horizontal line) we define the low threshold (in p.e.) for the considered different epochs of the mission as 5.2, 6.8, 7.8, 8.8 p.e. for BoL, after the first year, after the second year, and at EoL, respectively.

Processing two random pixel coincidences will amount to a total of 250~$\mu$s per second, while processing the 240~Hz expected trigger rate from single and two pixels configurations will reach $\sim$7.4~ms which corresponds to a dead time below 1\%.

\section {Conclusions}
Terzina is a pathfinder mission to demonstrate Cherenkov detection from a space platform of middle size enabling the possibility to perform neutrino astronomy from space and measuring UHECRs. Its results will be useful information for the planned POEMMA mission~\cite{pub:poemma}.
Rapidly evolving progress on the development of the Terzina telescope and its SiPM camera, including the full simulation chain, the amplification stage, and the ASIC, is reported. 
A preliminary estimation of  the trigger rate at a given threshold and readout pad size has been assessed for different mission lifetimes.
The radiation dose accumulated by the SiPM and electronics mainly caused by electrons, secondary gammas, and protons has been estimated as well as the SiPM power consumption. 
An indication of Terzina aperture has been provided.


\section*{Acknowledgements}

NUSES is a joint project of the Gran Sasso Science Institute and Thales Alenia Space Italia, funded by the Italian Government (CIPE  n. 20/2019 ), by the Italian Minister of Economic Development and the Abruzzo Region (MISE  n. F/130087/00/X38), by the Italian Space Agency (ASI n. 15/2022) and by the Swiss National Foundation (SNF grant n.~178918). 



\begin{thebibliography}{}
\bibitem{ref:tocome}
I.~De~Mitri for the NUSES Collaboration, J. Phys.: Conf. Ser. 2429 (2023) 012007.

\bibitem{Adriano:2022}
A.~Di~Giovanni and M. di Santo for the NUSES Collaboration, POS 414 (2022) 354, 
URL: https://pos.sissa.it/414/354.

\bibitem{Cummings:2020ycz}
A.~L.~Cummings, R.~Aloisio and J.~F.~Krizmanic,
``Modeling of the Tau and Muon Neutrino-induced Optical Cherenkov Signals from Upward-moving Extensive Air Showers,''
Phys. Rev. D \textbf{103} (2021) no.4, 043017
[arXiv:2011.09869].

\bibitem{pub:main_pub}
  A.~L.~Cummings, R.~Aloisio, J. Eser and J.~F.~Krizmanic, ``Modeling the Optical Cherenkov Signals by Cosmic Ray Extensive Air Showers Directly Observed from Sub-Orbital and Orbital Altitudes,'' Phys. Rev. D104 (2021) 063029.

\bibitem{pub:poemma}
A.~V.~Olinto \textit{et al.} [The POEMMA Collaboration],
``The POEMMA (Probe of Extreme Multi-Messenger Astrophysics) observatory,''
JCAP \textbf{06} (2021), 007,
[arXiv:2012.07945].


\bibitem{pub:poemma_pub}
 J. Krizmanic, et al. "POEMMA: Probe of extreme multi-messenger astrophysics." EPJ Web of Conferences. Vol. \textbf{210}. EDP Sciences, 2019.

 

\bibitem{pub:NuV-HD}
  A. Gola et al. "NUV-Sensitive Silicon Photomultiplier Technologies Developed at Fondazione Bruno Kessler", Sensor \textbf{19} (2019) 308. 



\bibitem{pub:SPENVIS}
  https://www.spenvis.oma.be/.
  

\bibitem{pub:geant4}
  S. Agostinelli et al., "GEANT4: A simulation toolkit", Nucl. Instrum. Meth. A506 (2003) 250–303.

\bibitem{pub:geant4_cite}
  http://geant4.web.cern.ch/geant4.


\bibitem{pub:easchersim}
 https://pypi.org/project/easchersim/1.1/, Austin Cummings BSD 3-Clause License.






\bibitem{PAO_2000}
J. Bellido et al.(Auger Collab.), Proceedings of Science(ICRC2017), 506 (2017).

\bibitem{icetop}
M.~G.~Aartsen et al., Phys. Rev. D 100 (2019) 082002.

\bibitem{pub:takahiko}
  M. Takahiko et al. "Suppression of the optical crosstalk in a multichannel silicon photomultiplier array" Opt. Express 29 (2021) 16914-16926. 
  

\bibitem{pub:NGB}
   W. Hofmann, et al. "Balloon-borne infrared telescope for absolute surface photometry of the night sky," Appl. Opt. \textbf{16} (1977) 3125-3130.  


\bibitem{pub:POLAR2}
  S. Mianowski et al. "Proton Irradiation of SiPM arrays for POLAR-2", Research Square (2022).
\end{thebibliography}
%

%
%


\section*{The NUSES Collaboration}
\small

\begin{sloppypar}\noindent

\noindent
R.$\,$Aloisio$^{1,2}$,
C.$\,$Altomare$^{3}$,
F.$\,$C.$\,$T.$\,$Barbato$^{1,2}$,
R.$\,$Battiston$^{4,5}$,
M.$\,$Bertaina$^{6,7}$,
E.$\,$Bissaldi$^{3,8}$,
D.$\,$Boncioli$^{2,9}$,
L.$\,$Burmistrov$^{10}$,
I.$\,$Cagnoli$^{1,2}$,
M.$\,$Casolino$^{13}$,
N.$\,$D'Ambrosio$^{2}$,
I.$\,$De Mitri$^{1,2}$,
G.$\,$De Robertis$^{3}$,
C.$\,$De$\,$Santis$^{11}$,
A.$\,$Di$\,$Giovanni$^{1,2}$,
A.$\,$Di$\,$Salvo$^{7}$,
M. Di Santo$^{1,2}$,
L. Di Venere$^{3}$,
M. Fernandez Alonso$^{1,2}$,
G. Fontanella$^{1,2}$,
P. Fusco$^{3,8}$,
S. Garbolino$^{7}$,
F. Gargano$^{3}$,
R. Aaron Giampaolo$^{1,7}$,
M. Giliberti$^{3,8}$,
F. Guarino$^{11,12}$,
M.$\,$Heller$^{10}$,
R.$\,$Iuppa$^{4,5}$,
A.$\,$Lega$^{4,5}$,
F. Licciulli$^{3}$,
F. Loparco$^{3,8}$,
L. Lorusso$^{3,8}$,
M. Mariotti$^{13,14}$,
M. N. Mazziotta$^{3}$,
M. Mese$^{11,12}$,
H. Miyamoto$^{1,7}$,
T. Montaruli$^{10}$,
A. Nagai$^{10}$,
R. Nicolaidis$^{4,5}$,
F. Nozzoli$^{4,5}$,
D. Orlandi$^{2}$,
G. Osteria$^{11}$,
P. A. Palmieri$^{6,7}$,
B.$\,$Panico$^{11,12}$,
G. Panzarini$^{3,8}$,
A. Parenti$^{1,2}$,
L. Perrone$^{15,16}$,
P.$\,$Picozza$^{17}$,
R. Pillera$^{3,8}$,
R. Rando$^{13,14}$,
M. Rinaldi$^{18}$,
A.$\,$Rivetti$^{7}$,
V. Rizi$^{2,9}$,
F. Salamida$^{2,9}$,
E. Santero Mormile$^{6}$,
V. Scherini$^{15,16}$,
V. Scotti$^{11,12}$,
D. Serini$^{3}$,
I. Siddique$^{1,2}$,
L.$\,$Silveri$^{1,2}$,
A. Smirnov$^{1,2}$,
R. Sparvoli$^{18}$,
S. Tedesco$^{7,19}$,
C.$\,$Trimarelli$^{10}$,
L. Wu$^{1,2}$,
P. Zuccon$^{4,5}$,
S. C. Zugravel$^{7,19}$.\\

\vskip 1.0cm\noindent
$^1$Gran Sasso Science Insttute (GSSI), Via Iacobucci 2, I-67100 L'Aquila,  Italy\\
$^2$Istituto Nazionale di Fisica Nucleare (INFN) - Laboratori Nazionali del Gran Sasso, I-67100 Assergi, L'Aquila, Italy\\
$^{3}$Istituto Nazionale di Fisica Nucleare, Sezione di Bari, via Orabona 4, I-70126 Bari, Italy\\
$^4$Dipartimento di Fisica, Università di Trento, via Sommarive 14 I-38123 Trento, Italy\\
$^5$Istituto Nazionale di Fisica Nucleare (INFN) - TIFPA, via Sommarive 14 I-38123 Trento, Italy\\
$^6$Dipartimento di Fisica, Università di Torino, Via P. Giuria, 1 I-10125 Torino, Italy\\
$^7$Istituto Nazionale di Fisica Nucleare (INFN) - Sezione di Torino, I-10125 Torino, Italy\\
$^{8}$Dipartimento di Fisica M. Merlin, dell’Università e del Politecnico di Bari, via Amendola 173, I-70126 Bari, Italy\\
$^9$Dipartimento di Scienze Fisiche e Chimiche, Università degli Studi di L'Aquila, I-67100 L'Aquila, Italy\\
$^{10}$Département de Physique Nuclèaire et Corpusculaire, Université de Genève, 1205 Genève, Switzerland\\
$^{11}$Istituto Nazionale di Fisica Nucleare, Sezione di Napoli, via Cintia, I-80126 Napoli, Italy\\
$^{12}$Dipartimento di Fisica E. Pancini dell'Università di Napoli Federico II, via Cintia, I-80126 Napoli, Italy\\
$^{13}$Università di Padova, I-35122 Padova, Italy\\
$^{14}$Istituto Nazionale di Fisica Nucleare (INFN) - Sezione di Padova, I-35131 Padova, Italy\\
$^{15}$Dipartimento di Matematica e Fisica ``E. De Giorgi", Università del Salento, Via per Arnesano, I-73100 Lecce, Italy\\
$^{16}$Istituto Nazionale di Fisica Nucleare - INFN - Sezione di Lecce, Via per Arnesano, I-73100 Lecce, Italy\\
$^{17}$RIKEN, 2-1 Hirosawa, Wako, Saitama, Japan\\
$^{18}$INFN Roma Tor Vergata, Dipartimento di Fisica, Universitá di Roma Tor Vergata, Roma, Italy\\
$^{19}$Department of Electrical, Electronics and Communications Engineering, Politecnico di Torino, Corso Duca degli Abruzzi 24, I-10129 Torino, Italy\\
%

\end{sloppypar}

\end{document}